\documentclass[%
aip,
amsmath,amssymb,
reprint,%
onecolumn,
]{revtex4-1}

\usepackage{graphicx}

\usepackage[utf8]{inputenc}
\usepackage[T1]{fontenc}
\usepackage{upgreek}

\usepackage{amsmath, amsthm, amssymb}

\usepackage{xcolor}
\newcommand{\revision}[1]{\textcolor{black}{#1}}

\begin{document} 
	%
	%
	%
	%
	%
	\title{Wafer-Scale Electroactive Nanoporous Silicon: Large and Fully Reversible Electrochemo-Mechanical Actuation in Aqueous Electrolytes}
	
	%
	%
	%
	%
	%
	%
	%
	%
	%
	%
	%
	\author{Manuel Brinker}
	\altaffiliation{manuel.brinker@tuhh.de}
	\affiliation{Institute for Materials and X-Ray Physics, Hamburg University of Technology, 21073 Hamburg, Germany}
	\affiliation{Centre for X-Ray and Nano Science CXNS, Deutsches Elektronen-Synchrotron DESY, 22607 Hamburg, Germany}
	\author{Patrick Huber}
	\altaffiliation{patrick.huber@tuhh.de }
	\affiliation{Institute for Materials and X-Ray Physics, Hamburg University of Technology, 21073 Hamburg, Germany}
	\affiliation{Centre for X-Ray and Nano Science CXNS, Deutsches Elektronen-Synchrotron DESY, 22607 Hamburg, Germany}
	\affiliation{Center for Hybrid Nanostructures CHyN, University of Hamburg, 22607 Hamburg, Germany}
	\keywords{nanoporous media, electrochemical characterization, actuorics, cyclic voltammetry, laser cantilever bending}
	%
	%
	%
	%
	%
	%
	\begin{abstract}
		Nanoporosity in silicon results in an interface-dominated mechanics, fluidics and photonics that are often superior to the ones of the bulk material. However, their active control, e.g. as a response to electronic stimuli, is challenging due to the absence of intrinsic piezoelectricity in the base material. Here, for large-scale nanoporous silicon cantilevers wetted by aqueous electrolytes, we show electrosorption-induced mechanical stress generation of up to $600\,\mathrm{kPa}$ that is reversible and adjustable at will by electrical potential variations of approximately $1\,\mathrm{V}$. Laser cantilever bending experiments in combination with in-operando cyclic voltammetry and step-coulombmetry allow us to quantitatively trace this large electro-actuation to the concerted action of 100 billions of parallel nanopores per square centimeter cross section and to determine the capacitive charge-stress coupling parameter upon ion ad- and desorption as well as the intimately related stress actuation dynamics for perchloric and isotonic saline solutions. A comparison with planar silicon surfaces reveals mechanistic insights on the observed electrocapillarity with respect to the importance of oxide formation and pore-wall roughness on the single-nanopore scale. The observation of robust electrochemo-mechanical actuation in a mainstream semiconductor with wafer-scale, self-organized nanoporosity opens up entirely novel opportunities for on-chip integrated stress generation and actuorics at exceptionally low operation voltages.
	\end{abstract}
	\maketitle
	\section{Introduction}
	\noindent Porous silicon has been attracting much attention since the discovery of self-organized porosity in a monolithic semiconductor.\cite{Lehmann1991,Canham1990,Lehmann1992,Stewart2000, Chiappini2010,Sailor2011, Huber2015} It provides a single-crystalline medium with anisotropic pores for the fundamental study of nanoconfinement on the structure and dynamics of matter~\cite{Gruener2008,Gor2015,Kondrashova2017, Huber2015,Vincent2016,Cencha2020}. Interface-dominated optical, electrical and thermal properties attract the attention of the applied sciences in fields ranging from in-vivo \cite{Stewart2000} and opto-electronics \cite{Nattestad2010} via biosensing \cite{Gu2013,Sailor2011} to energy storage \cite{Westover2014,Jia2020} and harvesting \cite{Valalaki2016}. Silicon as a raw material constitutes one of the elements most commonly found on earth and is available in unique qualities. An integration of wafer-scale porous silicon into electrical circuit dsigns already existing could reasonably be achieved, as semiconductor devices devices are predominantly fabricated out of bulk silicon.\cite{Sailor2011} Furthermore, porous silicon has been found to be bio-compatible\cite{Canham1995} and a lot of functionalization schemes exist, which are using subseqeuent treatments to change, extend or enhance its properties by incorporating additional functional materials~\cite{Stewart2000,Li2014,Kityk2014,TzurBalter2015,Brinker2020,Zhang2021LiquidBasedMaterials} or, as recently shown, by laser writing directly in pore space \cite{Ocier2020}. In particular, functionalization with liquids offers a plethora of opportunities to tune the properties and to create adaptive hybrids, where the soft, dynamic filling, affected by confinement, provides novel functionalities, but the rigid, semiconducting scaffold structure provides mechanical robustness on the macroscopic scale.\cite{Canham2015,Zhang2021LiquidBasedMaterials}\\
	An aspect that has not yet prompted significant research is the mechanics and particularly a functionalization of porous silicon as an actuator material. Porous silicon among other, different porous media has been investigated with a focus on humidity and gas sorption-induced actuation.\cite{Zhao2014,Ganser2016,Vanopdenbosch2016}  Also liquid adsorption-induced deformation of mesoporous silicon has been explored\cite{Grosman2015, Gor2015,Gor2017,Rolley2017} to study the mechanical properties of mesoporous silicon or its functionalization with artificial muscle molecules.\cite{Brinker2020} Another promising direction of research could be actuation mechanisms due to an electrochemical process within the material or on its surface.\cite{Kong2014,Liang2011,Liu2019} Given the absence of intrinsic piezoelectrocity in silicon, an actuator functionality of silicon has been so far achieved either by piezo-ceramic thin film on-silicon coatings or epitaxial heterostructures, most prominently in the context of nano- and micromechanical systems (NEMS/MEMS).\cite{Baek2011}\\
	We follow here the strategy to integrate mechanical actuation into porous silicon by exploiting electrical potential-induced surface stress at the internal pore surfaces, as it has been extensively studied in the domain of porous metal actuation.\cite{Jin2010,Seker2009} The interface between a conductor immersed in an electrolyte solution, i.e. an ionic conductor, can be electrically polarized by the formation of a Helmholtz double layer.\cite{Helmholtz1853} Thus, the accumulated charge ($q$ - surface charge density per area) on the electrode surface leads to a change in surface stress $f$. So,
	\begin{equation}
		\delta f=\varsigma\delta q,
		\label{Eq_surface-stress}
	\end{equation} 
	where $\varsigma=\mathrm{d}f/\mathrm{d}q\rvert_e$ denotes the electrocapillarity coupling parameter and $e$ the tangential strain per area. \revision{Essentially, the accumulation of charge carriers influences the bonds at the surface in the in-plane direction and normal to it and a charge reorganization ensues which leads electrostatic (Hellman-Feynman) forces on the surface ions.\cite{Weigend2006} Ab-initio calculations confirm, that an electrochemo-mechanical coupling effect can generally also be observed in silicon.\cite{Hoppe2014}} The change in surface stress in itself can induce an actuation effect of a non-porous, bulk material, as already demonstrated for a clean gold surface.\cite{Smetanin2008} Platinum is another metal that has shown a similar electrochemical actuation due to an evolving surface tension by the accumulation of charge carriers.\cite{Raiteri1995} In this study, the fundamental relation of surface stress and accumulated charge and the resulting potential driven actuation is demonstrated for the first time for porous and planar, non-porous silicon surfaces.\\
	For bulk materials this actuation effect is comparatively small, due to their small surface to volume ratio. For nanoporous solids with a significant internal surface area though, a change in surface stress can induce a considerable strain of the entire porous solid.\cite{Weissmuller2003} In thermodynamics a Maxwell relation then equates the charge-response of $f$ to the potential-response on tangential strain $e$,\cite{Weissmuller2013,Stenner2016}
	\begin{equation}
		\varsigma=\frac{\mathrm{d}f}{\mathrm{d}q}\biggr\rvert_{e}=\frac{\mathrm{d}E}{\mathrm{d}e}\biggr\rvert_{q}.
		\label{Eq_Maxwell}
	\end{equation} 
	The microscopic description of these phenomena can be employed on macroscopic porous materials, by considering it as an effective medium and normalizing to volume $V$-designated quantities. Thus, an effective strain-charge actuation coefficient is introduced, as $A^{*}=\mathrm{d}\varepsilon/\mathrm{d}q_\mathrm{V}\rvert_T$, where $T$ is the load per cross-sectional area. By means of $A^{*}$ the performance of the porous silicon's electrochemical actuation is assessed.\\
	Therefore, the porous silicon electrode has to comply with the precondition that it exhibits polarizability. It means that a finite potential range in an electrochemical measurement exists where electrolyte charge carriers exclusively accumulate on the electrode surface in an electric double layer.\cite{Butt2003} In contrast, Faradaic reactions and thus currents occur in an electrochemical reaction of the involved species.\cite{Weissmuller2013} The resulting effect of the double layer on the surface stress may vary. Metals exhibit a high conductivity and therefore charge carriers of the opposite sign accumulate at the metals surface in an $\buildrel _\circ \over {\mathrm{A}}$ thin layer to screen the material from the electrical field induced by the double layer.\cite{Weissmuller2003} By contrast, the screening layer within a semi-conductor can grow to several $100\,\buildrel _\circ \over {\mathrm{A}}$.\cite{Zhang2007}. Moreover, the respective porous structure of nanoporous materials may differ greatly. The noble metal based nanoporous gold has an entirely different porous structure than porous silicon investigated in this study. The structure of nanoporous gold is formed by so called ligaments, which are interconnected in an isotropically oriented network. The here investigated porous silicon on the other hand has a highly anisotropic porous structure, which consists of parallel pores, orthogonally oriented to the silicon surface.\\
	Overall, here we combine surface stress induced electrochemical actuation in planar bulk- and porous silicon and the assessment of their respective properties.
	%
	%
	%
	%
	%
	%
	\section{Results}
	\subsection{Electrochemo-Mechanical Actuation of Nanoporous Silicon}
	\label{sec_psi}
\begin{figure}
		\centering
		\includegraphics{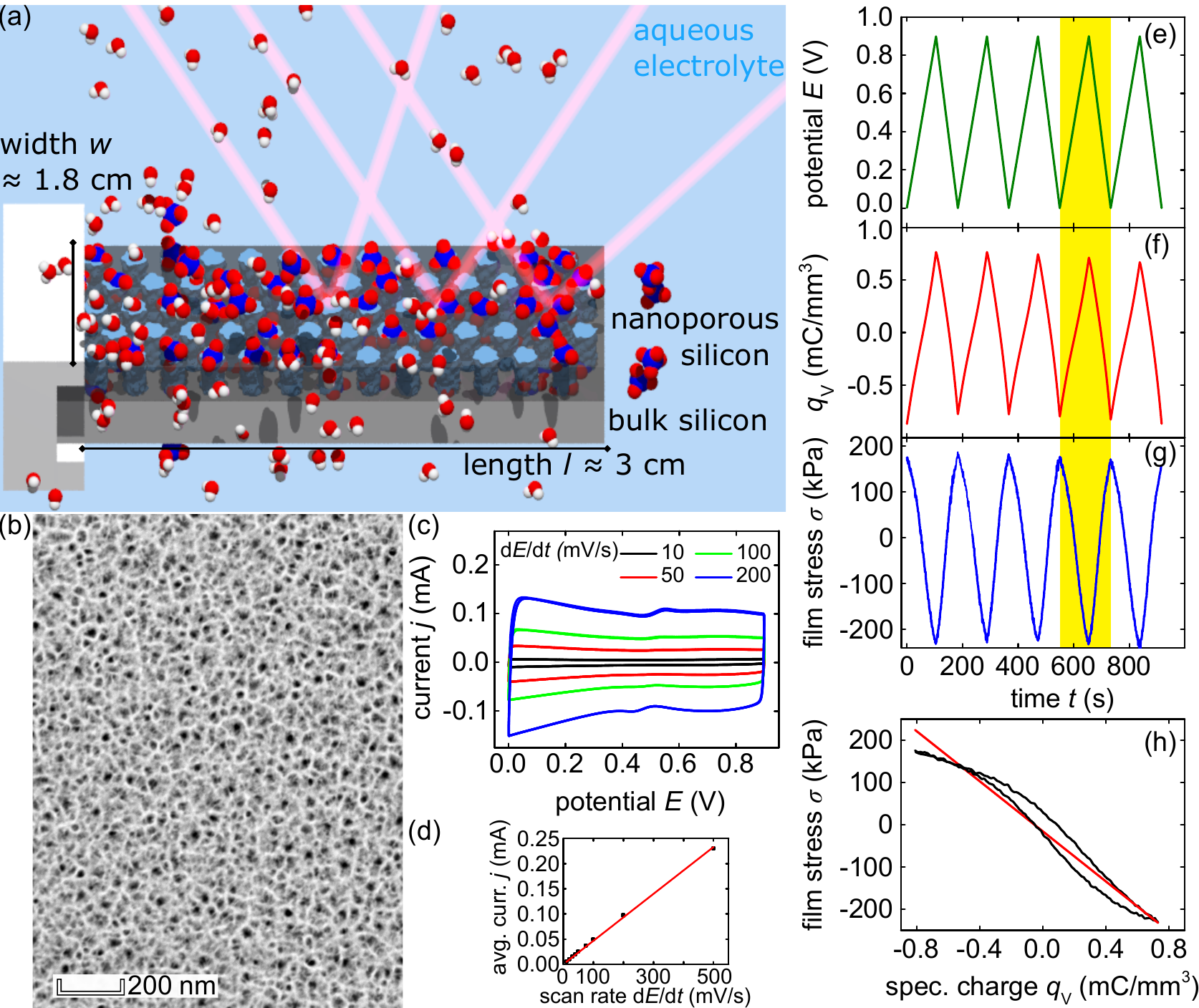}
		\caption{\textbf{In-operando laser cantilever bending experiments on electrosorption-induced actuation of nanoporous silicon immersed in aqueous electrolyte.} (a) Schematic illustration of the in-situ cantilever bending measurement on a porous silicon sample (light blue pores) with a bulk silicon layer underneath (light gray) in an aqueous (red, white molecules) perchloric acid (blue, red molecules) electrolyte solution. The length and width of the surface area in contact with the electrolyte solution are $l=2.990\,\mathrm{cm}$ and a width of $w=1.796\,\mathrm{cm}$. (b) TEM micrograph of p-doped porous silicon. (c) The graph depicts exemplary cyclic voltammetry measurements of a porous silicon sample in $1\,\mathrm{mol\, l^{-1}}$ $\mathrm{HClO_4}$ electrolyte solution. The current $j$ is plotted against the applied potential $E$ in the range of $0.0\,-\,0.9\,\mathrm{V}$, measured versus SHE. The potential scan rate is increased from $10\,\mathrm{mVs^{-1}}$ to $200\,\mathrm{mVs^{-1}}$. (d) The graph depicts values for the current $j$ from the cyclic voltammetry measurements depicted in (c), averaged in the potential range of $0.35\,-\,0.45\,\mathrm{V}$. The values are plotted against increasing potential scan rates $\mathrm{d}E/\mathrm{d}t$ from $10\,\mathrm{mVs^{-1}}$ to $500\,\mathrm{mVs^{-1}}$. The red linear regression line yields the capacitance $c=0.466\pm0.004\,\mathrm{m F}$ as the slope. On the right side the respective curves of the electrochemical actuation measurement with a scan rate of $10\,\mathrm{mVs^{-1}}$, are shown (e) 5 representative potential cycles $E$, (f) the resulting volume specific charge $q_\mathrm{V}$ and (g) the introduced change in film stress $\sigma$ in the porous silicon layer. (h) Change in film stress $\sigma$ versus deposited volume specific charge $q_\mathrm{V}$ averaged over the 5 cycles and the linear fit to determine the stress charge coupling parameter $\xi$.}
		\label{Fig_Actu_pSi_perc}
	\end{figure}

\begin{figure}
		\centering
		\includegraphics{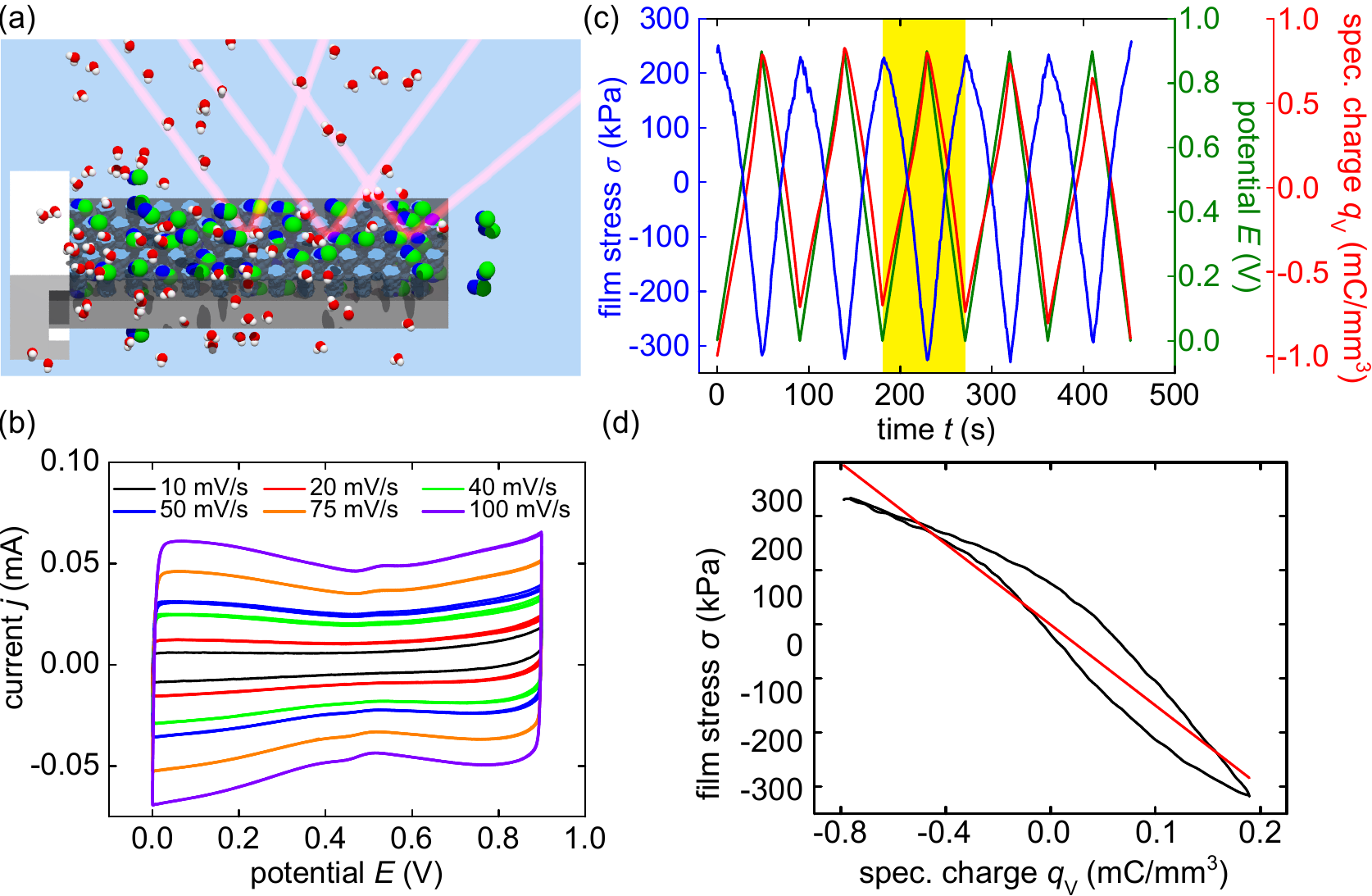}
		\caption{\textbf{In-operando laser cantilever bending experiments on electrosorption-induced actuation of nanoporous silicon immersed in isotonic saline solution.} (a) Schematic illustration of the cantilever bending measurement on a porous silicon sample in aqueous (red, white molecules) NaCl (green, blue molecules) electrolyte solution. (b) CV measurements in an isotonic saline solution ($154\,\mathrm{mmol\, l^{-1}}$ NaCl aq.) with different scan rates in the potential region of $0.0$ to $0.9\,\mathrm{V}$. The ascertained capacitance is $c=0.473 \pm 0.003\,\mathrm{m F}$. (c) Depicted are the respective curves of the electrochemical actuation measurement with a scan rate of $10\,\mathrm{mVs^{-1}}$ in the same NaCl aqueous electrolyte solution versus time, as 5 representative potential cycles of $E$ (green), $q_\mathrm{V}$ (red) and $\sigma$ (blue). The film stress $\sigma$ averaged over 5 cycles versus deposited volume specific charge $q_\mathrm{V}$ yields $\xi=-374 \pm 2 \,\mathrm{mV}$ via the respective linear fit. (d) Averaged film stress $\sigma$ versus deposited volume specific charge $q_\mathrm{V}$ and the linear fit to determine the stress charge coupling parameter $\xi= -374 \pm 2 \,\mathrm{mV}$.}
		\label{Fig_Actu_pSi_iso}
	\end{figure}

	The first part of the study will deal with the electrochemical and actuation properties of porous silicon. A transmission electron microscopy (TEM) micrograph of a cross section of the resulting p-doped porous silicon is depicted in Figure \ref{Fig_Actu_pSi_perc}(b).
	The fabrication process is described in detail in the experimental section. The TEM cross section shows that the pore network resembles a randomized honeycomb structure, as discussed in more detail in~\cite{Brinker2020}. A scanning electron microscope (SEM) micrograph of the 
	sample material is shown in Figure S1 in the supplementary materials section. Another SEM micrograph of the entire porous sample profile gives the thickness of the porous silicon layer as $h_\mathrm{f}=630\,\mathrm{nm}$. The total thickness of the sample amounts to $110\pm 2\,\mathrm{\upmu m}$. Thus, the remaining bulk silicon layer has a thickness of $109.37\,\mathrm{\upmu m}$. The volume of the porous silicon layer is then $V_\mathrm{pSi}=l\cdot w \cdot h_\mathrm{f}=0.338\,\mathrm{mm^3}$, where $l$ and $w$ denote the length and width of the sample, further defined in the experimental section.\\
	A nitrogen sorption isotherme, shown in Figure S2 in the supplementary materials, results in a mean pore radius of $r=3.36\,\mathrm{nm}$, an internal surface area $A_\mathrm{por}=4.003\,\mathrm{m^2}$ and a porous volume of $V_\mathrm{por}=0.009\,\mathrm{cm^3}$, which corresponds to a porosity of $54\,\mathrm{\%}$.\\
	The porous silicon sample is installed in the cantilever bending setup and immersed in 1 \textsc{m} $\mathrm{HClO_4}$ electrolyte solution, as sketched in Figure \ref{Fig_Actu_pSi_perc}(a). An initial electrochemical characterization by cyclic voltammetry (CV) measurements is conducted to examine, if porous silicon can be used as a polarizable electrode. The respective measurement is shown in Figure S3 in the supplementary materials. It becomes apparent, that the as fabricated porous silicon is subjected to an oxidation process. Thereby, applying a potential leads to a gradually decreasing anodic oxidation of the porous silicon walls.\cite{Bsiesy1991,Gaspard1987,Zhang2007} The electrochemically irreversible character of the oxidation is further discussed in the supplementary materials. The oxidative currents have to be drawn to a close to foreground the capacitive characteristics of the material. Therefore, a constant potential of $1.2\,\mathrm{V}$ is applied for 20 hours. Following this preparation step, a slightly constrained potential range of $0\,\mathrm{V}$ to $0.9\,\mathrm{V}$ will be explored in the following with the accompanying cantilever bending measurements.\\
	Figure \ref{Fig_Actu_pSi_perc}(c) shows cyclic voltammetry measurements in the specific potential range of $0.0\,\mathrm{V}$ to $0.9\,\mathrm{V}$ with different scan rates from $10\,\mathrm{mVs^{-1}}$ to $200\,\mathrm{mVs^{-1}}$ for the oxidized porous silicon sample. Increasing the potential $E$ from the lower vertex, causes the current $j$ to increase instantly to a near constant value. Vice versa, the current quickly decreases to its negative counterpart, when the sweep direction is changed at the upper vertex. Thus, the course resembles a square around zero. This behavior holds true for the depicted courses for all scan rates $\mathrm{d}E/\mathrm{d}t$, albeit larger scan rates from  $100\,\mathrm{mVs^{-1}}$ on exhibit a slight decrease from the lower vertex of the CV. In particular, neither an increasingly extended time until constant current, nor a linearly increasing current, instead of a constant current, is apparent. By this means, the sample seemingly shows no sign of a diffusion limited kinetic of the charge carrier up to $200\,\mathrm{mVs^{-1}}$.\cite{Hamann2007} Thus, the charge carriers move to the working electrode, i.e. the sample, in a constant flow to counterbalance an increasing potential. The actual potential is not affecting the amplitude of the current though, just the sweep direction and scan rate. The capacitance $c$ of the sample determines the constant current value. The rate of the potential change $\mathrm{d}E/\mathrm{d}t$ adjusts the current according to the capacitance $c=\mathrm{d}j\,\mathrm{d}t/\mathrm{d}E$.\cite{Roschning2019} Hence, it is possible to extract the sample's capacitance $c$ by a linear regression of $j$ versus an increasing scan rate. The absolute value of the current for each CV for both sweep directions is averaged in the range of $0.35-0.45\,\mathrm{V}$ and plotted versus the respective scan rate. The resulting plot is depicted in Figure \ref{Fig_Actu_pSi_perc}(d). A linear dependence is visible up to scan rates of $500\,\mathrm{mVs^{-1}}$. The capacitance is then obtained by a linear fit to the data and amounts to $c=0.466\pm0.004\,\mathrm{m F}$.\\
	The overall electrochemical analysis is evidence of the stable, capacitive features of the oxidized porous silicon electrode and thus the material can be considered as polarizable. For a complete electrochemical characterization the determination of the potential of zero charge would be ideal, albeit challenging.\cite{Dickertmann1976,Hamelin1985,Smetanin2008}\\
	Recording the change in film stress $\sigma$ during a CV measurement allows a detailed characterization of the electrochemical actuation of the sample. The respective measurements are depicted in Figure \ref{Fig_Actu_pSi_perc}(e)-(h). While the applied potential is reversibly changed from $0.0\,\mathrm{V}$ to $0.9\,\mathrm{V}$ with a scan rate of $10\,\mathrm{mVs^{-1}}$ (see Figure \ref{Fig_Actu_pSi_perc}(e)), the current is measured and hence the transferred charge can be quantified. It is normalized to the sample volume $V_\mathrm{pSi}$ and henceforth referred to as the volume specific charge $q_\mathrm{V}$, depicted in Figure \ref{Fig_Actu_pSi_perc}(f). The course of $q_\mathrm{V}$ clearly linearly coincides with the applied potential $E$. Thus, increasing the potential leads to an increased accumulation of charge carriers in the electrochemical double layer and vice versa, a decreasing potential expels charge carriers. Similarly, the change in film stress $\sigma$ linearly follows the potential and the specific charge but with a negative proportionality, see Figure \ref{Fig_Actu_pSi_perc}(g). So, an increasing potential leads to an accumulation of charge carriers and is then responsible for a decrease in film stress $\sigma$, which is caused by a contraction of the porous silicon layer. Subsequently, a decreasing potential deprives the double layer of charge carriers and results in an expansion of the sample and thus an increase in film stress $\sigma$. The amplitude of $\sigma$ exhibits no sign of a decrease and reversibly moves from a level of $-210\,\mathrm{kPa}$ to $170\,\mathrm{kPa}$ from cycle to cycle. Thus, our experiments demonstrate a robust and reversible electrosorption-induced actuation in porous silicon.\\
	Additionally, by averaging $\sigma$ and $q_\mathrm{V}$ over the 5 cycles, further details can be inferred about the electrochemical actuation. The average yields a peak-to-peak amplitude of the film stress $\sigma_\mathrm{avg}=406.27\pm0.3\,\mathrm{kPa}$ and of the charge $q_\mathrm{V_{avg}}=1.54\pm0.04\,\mathrm{mCmm^{-3}}$. Figure \ref{Fig_Actu_pSi_perc}(h) shows the average film stress $\sigma$ plotted against the accumulated specific charge $q_\mathrm{V}$ in the associated potential range of $0.0\,-\, 0.9\,\mathrm{V}$. It is visible, that the relation between the two is not entirely linear, but has a slightly smaller slope in the beginning from approximately $-0.8$ to $-0.4\,\mathrm{mCmm^{-3}}$. Moreover, it shows a hysteresis from $-0.4\,\mathrm{mCmm^{-3}}$ on. This behavior is potentially the result of retardation in the stress response on the charge accumulation. Nonetheless, a linear fit to the data yields the corresponding stress charge coupling parameter $\xi = -296 \pm 1 \,\mathrm{mV}$. $\xi$ is a key materials parameter to asses the electrochemo-mechanical coupling that drives the actuation of porous silicon.\\
	To explore the potential bio-medical application of this process, analogous measurements are repeated with an aqueous electrolyte solution of isotonic sodium chloride, as omnipresent in (bio-)medical environments, see Figure \ref{Fig_Actu_pSi_iso}(a). 
	Section (b) shows the respective CVs with scan rates from $10\,-\,100\,\mathrm{mVs^{-1}}$. The CV measurement are conducted in the same potential range, exhibiting the characteristics of a capacitive regime as well. Moreover, the same small Redox-like peaks in the up- and down sweep are visible with the NaCl electrolyte solution as well. Through the fit of the averaged absolute current of both sweep directions between $0.35\,-\,0.45\,\mathrm{V}$ versus the scan rate, the capacitance of the sample is determined. It yields a value of $c=0.473 \pm 0.003\,\mathrm{m F}$, which is in excellent agreement with the value obtained with $\mathrm{HClO_4}$ as the electrolyte. Figure S4 in the supporting information shows the fit for $c$.\\
	In Figure \ref{Fig_Actu_pSi_iso}(c) the electrochemical actuation measurement in the corresponding potential range with the NaCl electrolyte solution and a scan rate of $10\,\mathrm{mVs^{-1}}$ is displayed. The film stress $\sigma$ shows the same characteristics as with $\mathrm{HClO_4}$. It follows the potential and the accumulated specific charge $q_\mathrm{V}$ linearly with a negative sign. Averaging the stress over the cycles results in an average value for $\sigma$ of $503.2 \pm 0.5\,\mathrm{kPa}$, while the accumulated volume specific charge $q_\mathrm{V}$ remains in the same range of $-0.8$ to $+0.8\,\mathrm{mm^3C^{-1}}$. The relation of $\sigma$ to $q_\mathrm{V}$ exhibits the same hysteretic behavior as for the $\mathrm{HClO_4}$ electrolyte. A fit to the averaged film stress versus the averaged specific charge yields the stress-charge coupling parameter with a value of $\xi = -374 \pm 2 \,\mathrm{mV}$. This means a larger absolute value than for $\mathrm{HClO_4}$. Figure \ref{Fig_Actu_pSi_iso}(d) shows the respective fit for $\xi$.\\ 
	To characterize the kinetics of the electrochemical actuation, the response of the film stress $\sigma$ and the specific charge $q_\mathrm{V}$ to a square potential from $0.0\,\mathrm{V}$ to $0.9\,\mathrm{V}$ is measured. The measurement is conducted in $\mathrm{HClO_4}$ as the electrolyte. 
	\begin{figure}
		\centering
		\includegraphics{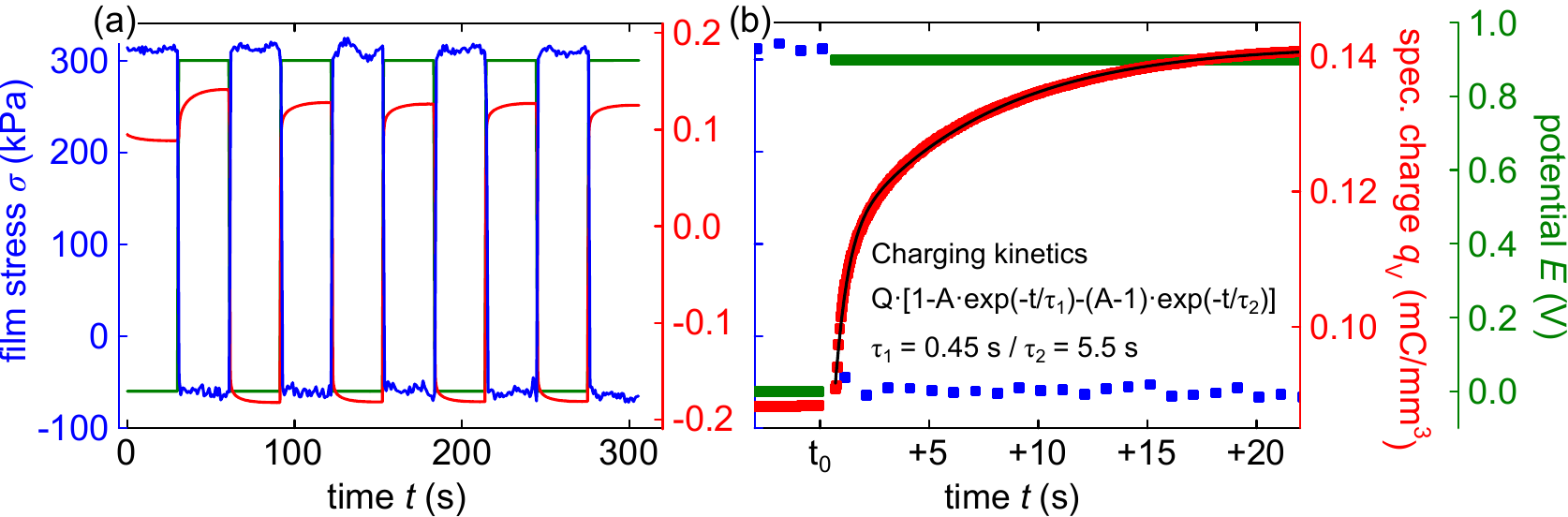}
		\caption{\textbf{Electroactuation kinetics of nanoporous silicon.} Step-coulombmetry is performed in $1\,\mathrm{mol\, l^{-1}}$ $\mathrm{HClO_4}$ electrolyte solution to determine the electrochemical actuation kinetics. The applied potential $E$ is changed in an instant from $0.0\,\mathrm{V}$ to $0.9\,\mathrm{V}$ and backwards thus resembling a square potential. The thereby incorporated volume specific charge $q_\mathrm{V}$ and the caused film stress $\sigma$ are measured versus time $t$. Whereas (a) shows the entire measurement, (b) depicts a close up of one potential increase from $0.0\,\mathrm{V}$ to $0.9\,\mathrm{V}$ at time $\mathrm{t}_0$. The signal of $q_\mathrm{V}$ is fitted by a sum of exponential functions (black).}
		\label{Fig_actu-perc-squ}
	\end{figure}
	The resulting graphs are depicted in Figure \ref{Fig_actu-perc-squ}(a) and (b). It is shown that $\sigma$ increases to its saturation level of approximately $+310\,\mathrm{kPa}$ with a decreasing potential step, while $q_\mathrm{V}$ decreases to a level of $-0.182\,\mathrm{mC mm^{-3}}$. Vice versa $\sigma$ and $q_\mathrm{V}$ reverse to approximately $-65\,\mathrm{kPa}$ and $0.125\,\mathrm{mC mm^{-3}}$ when the potential is increased to $0.9\,\mathrm{V}$.\\
	The charging and discharging of $q_\mathrm{V}$ upon voltage increase and decrease is fitted by the sum of two exponential functions in the form of $Q_\mathrm{ampl}\cdot\left[1-A\cdot\exp(-t/\tau_1)-(1-A)\cdot \exp(-t/\tau_2)\right]$, where $Q_\mathrm{ampl}$ and $A$ denote amplitude parameters and $\tau_1$ and $\tau_2$ characteristic time constants of the two respective exponential functions. This approach has been found to approximate the charging characteristics of supercapacitor materials, in response to a square potential, well. \cite{Breitsprecher2020,Pean2014} The fit is depicted on an exemplary cycle in Figure \ref{Fig_actu-perc-squ}(b). The respective obtained quantities, averaged over the 5 in- and decreasing potential steps, amount to, $\tau_{1,q_\mathrm{V},\mathrm{decr}} = 0.072 \pm 0.005\,\mathrm{s}$ for the fast volume specific charge decrease, $\tau_{2,q_\mathrm{V},\mathrm{decr}} = 3.3 \pm 0.5\,\mathrm{s}$ for the slow volume specific charge decrease, $\tau_{1,q_\mathrm{V},\mathrm{incr}} = 0.46 \pm 0.02\,\mathrm{s}$ for the fast volume specific charge increase and $\tau_{2,q_\mathrm{V},\mathrm{incr}} = 5.7 \pm 0.1\,\mathrm{s}$ for the slow volume specific charge increase. Thus, the sample is charged and discharged on two different time scales, whereby one is noticeably quicker.
	The change in film stress upon the instant potential increase occurs too fast for the setup's time resolution of $1\,\mathrm{s}$, cf. the close up in Figure \ref{Fig_actu-perc-squ}(b). Hence, the electro-mechanical response is faster than the overall charging kinetics, on a time scale obviously smaller than $1\,\mathrm{s}$. Consequently, the stress response kinetics seem to be connected with the faster charge movement process. Hence, one may speculate that the quick mechanical response is due to the fast formation of the charge layer closest to the pore wall-electrolyte interface, i.e. the Stern layer, whereas the remaining, slower build-up of the rather diffuse, Gouy-Chapman counter-ion layer contributes negligibly to the electrocapillarity response.\\
	An estimate for the timescale of charging $t_\mathrm{c}$ can be made by assuming diffusion of ions in an ideal cylindrical pore,\cite{Biesheuvel2011,Gupta2020}
	\begin{equation}
		\label{Eq_tc}
		t_\mathrm{c}=\frac{2\lambda}{a}\frac{\ell_\mathrm{pore}^2}{D},
	\end{equation}
	where $a=3.36\,\mathrm{nm}$ and $\ell_\mathrm{pore}=630\,\mathrm{nm}$ denotes the pore diameter and length respectively. $\lambda$ is the Debye length and is given by $\lambda=0.304\,\mathrm{nm}/c_0=0.304\,\mathrm{nm}$ for a monovalent salt diluted in water at $298\,\mathrm{K}$, where $c_0=1$ in $\,\mathrm{mol\, l^{-1}}$ gives the concentration of perchlorate.\cite{Butt2003} The diffusion coefficient is $D\approx2\cdot 10^{-9}\,\mathrm{m^2s^{-1}}$.\cite{Heil1995} Thus, $t_\mathrm{c}\approx36\,\mathrm{\upmu s}$ is much smaller than the measured charging time scales. Note however, that the silicon oxide in contact with aqueous eletrolytes is hydroxylated. For such hydrophilic surfaces both experiments \cite{Goertz2007,Gruener2009,Ortiz2013,Jani2021} and computer simulations \cite{Sendner2009, Schlaich2016WaterConfinement, Schlaich2017HydrationFriction} suggest the formation of highly viscous interfacial water layers of approximately 1 nm thickness. In these layers the corresponding water self-diffusivities is expected to exponentially increase towards the solid wall to values up to two orders of magnitude larger than in bulk water.\cite{Schlaich2017HydrationFriction} Thus, the reduced mobility of interfacial water molecules and the corresponding reduction in the ion self-diffusivity may explain at least partially the extremely slowed down charging kinetics observed here compared to the idealistic considerations, in particular the assumption of bulk self-diffusivities for the confined perchlorate ions. The partially dentritic and large pore surface roughness, see also the discussion below, may increase the effective diffusion length and thus additionally contribute to a reduced charging dynamics.\\
	Still, it is worthwhile to mention that the overall charging and stress response kinetics is much faster than observed for a hybrid material of porous silicon filled by an electrically conductive polymer.\cite{Brinker2020} There, the ions have to penetrate the confined polymer during the charging process, which leads to longer characteristic times around $15\,\mathrm{s}$ for the charging and $3\,\mathrm{s}$ for the mechanical response, respectively. Thus, the actuation scheme explored here is superior in terms of actuation kinetics.\\
	%
	%
	%
	%
	%
	%
	%
	\begin{figure}
		\centering
		\includegraphics{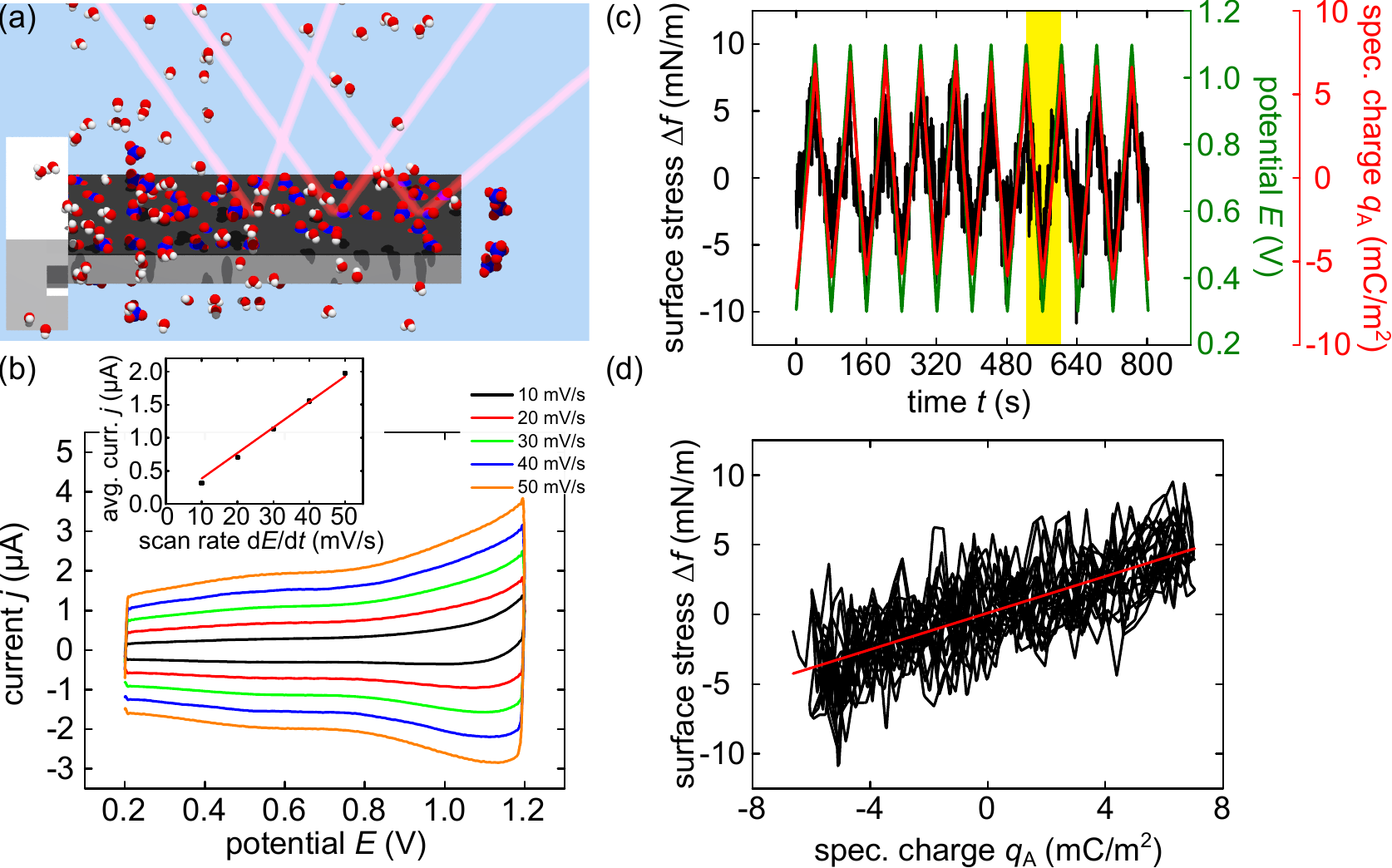}
		\caption{\textbf{Electrochemistry and in-situ laser cantilever bending experiments on bulk silicon immersed in aqueous electrolytes.} (a) Schematic depiction of the in-situ cantilever bending measurement on a bulk silicon sample (light gray) with a silicon oxide layer on top (dark gray) in an aqueous (red, white molecules) perchloric acid (blue, red molecules) electrolyte solution. The sample is clamped on the left by PTFE. The sample that is in contact with the electrolyte solution has a length of $4.28\,\mathrm{cm}$ and a width of $1.49\,\mathrm{cm}$. (b) The graph shows exemplary CV measurements of a bulk silicon sample in $1\,\mathrm{mol\, l^{-1}}$ $\mathrm{HClO_4}$ electrolyte. The current $j$ is plotted against the applied potential $E$. Five measurements with scan rates from $10\,\mathrm{mVs^{-1}}$ to $50\,\mathrm{mVs^{-1}}$ are shown. The inset depicts values for the current $j$, averaged in the potential range of $0.6\,-\,0.8\,\mathrm{V}$, plotted against increasing potential scan rates. The red line gives a linear regression of the data and yields a capacitance of $c=38.51\pm0.08\,\mathrm{\upmu F}$. (c) Measurement of the change in surface stress $\Delta f$ in a $1\,\mathrm{mol\, l^{-1}}$ $\mathrm{HClO_4}$ solution under a changing potential $E$ with a scan rate of $20\,\mathrm{mVs^{-1}}$ in the potential region of $0.3\,-\,1.1\,\mathrm{V}$. The surface area specific capacitance $q_\mathrm{A}$ is depicted as well. (d) Change in surface stress $\Delta f$ versus the specific charge $q_\mathrm{A}$. To independently determine the surface stress - charge coefficient $\varsigma$ a linear regression is used. It yields $\varsigma=+0.657\pm0.007\,\mathrm{V}$.}
		\label{Fig_Actu-bulkSi}
	\end{figure}
	\subsection{Electrochemo-Mechanical Actuation at Bulk Silicon Surfaces}
	\label{sec_bulk}
	To gain mechanistic and detailed quantitative insights on the actuation of porous silicon, also experiments on the planar, non-porous silicon surface are performed. A piece of bulk silicon is immersed into an electrolyte solution of 1 \textsc{m} $\mathrm{HClO_4}$. Thereby, the area of the sample in contact with the electrolyte amounts to $24.346\,\mathrm{cm^2}$. To render the measurements performed on bulk silicon comparable to porous silicon, the samples are treated in the same manner and the surface of bulk silicon is oxidized by applying a constant potential of $1.2\,\mathrm{V}$ for 15 hours. The reasons for applying a constant potential to the sample beforehand lie in the oxidative nature of the silicon surface as discussed in great detail above and in the supplementary materials. The recorded current is decreasing to $92\,\mathrm{nA}$. Subsequently CV measurements are performed, depicted in Figure \ref{Fig_Actu-bulkSi}(b). The potential range is $0.2\,-\,1.2\,\mathrm{V}$ with scan rates from $10-50\,\mathrm{mVs^{-1}}$. A characteristic capacitive charging is apparent for all depicted scan rates. Albeit for larger potentials from $0.8\,\mathrm{V}$ on for the fastest scan rate of $50\,\mathrm{mVs^{-1}}$ a slight increase towards the upper vertex of the CV becomes apparent. No distinct oxidation- or reduction peaks on the up- and down-sweep respectively are noticeable.\\
	The plot to determine the capacitance $c$ is depicted in the inset of Figure \ref{Fig_Actu-bulkSi}(b). The constant current for each scan rate is gained by averaging the absolute current values between $0.6\,-\,0.8\,\mathrm{V}$ for both sweep directions. A linear dependence is visible up to scan rates of $50\,\mathrm{mVs^{-1}}$. The linear regression yields a capacitance of $c=38.51\pm0.08\,\mathrm{\upmu F}$. Additionally, the roughness of the oxidized bulk silicon surface has been determined with an atomic force microscope. The roughness increases the apparent surface by a factor of $1.00075$. Therefore, the bulk silicon sample has a surface specific capacitance of $c^{*}=1.579\pm0.003\,\mathrm{\upmu Fcm^{-2}}$.\\
	Next, the measurement of a bulk silicon sample in the cantilever bending setup is analyzed. CV measurements with scan rates of $10\,,20\,,30\,,40\,,50$ and $100\,\mathrm{mVs^{-1}}$ are performed. The resulting measurements all exhibit the same capacitive characteristics already seen for the bulk silicon sample in Figure \ref{Fig_Actu-bulkSi}(b), and are shown in the supporting information in Figure S5(a). The linear regression of the averaged currents plotted versus the respective scan rate, as shown in Figure S5(b), yields a capacitance of $c=10.07\pm0.08\,\mathrm{\upmu F}$. Therefore, the surface area, that is in contact with the electrolyte solution inside the cell of the cantilever bending setup amounts to $c/c^{*}=10.07\,\mathrm{\upmu F}/1.579\,\mathrm{\upmu Fcm^{-2}}=6.38\pm0.05\,\mathrm{cm^2}$. The respective measurement of the surface stress for the CV measurement with a scan rate of $20\,\mathrm{mVs^{-1}}$ is depicted in Figure \ref{Fig_Actu-bulkSi}. The charge accumulation related curvature change $\Delta \kappa$ is already converted to a change in surface stress $\Delta f$ according to Equation \ref{Eq_Stoney}. Figure \ref{Fig_Actu-bulkSi}(c) shows an identifiable surface stress change of the bulk silicon sample. Hence, to highlight this point, the laser cantilever experiments provide the sensitivity to measure an electrochemical actuation in a bulk silicon sample. The measurement exhibits a change in surface stress $\Delta f$ that is well in phase with the potential $E$ and the accumulated charge per sample surface $q_\mathrm{A}$. The plot of $\Delta f$ versus $q_\mathrm{A}$ in (d) further emphasizes this linear relation. Here, a linear regression of the change in surface stress $\Delta f$ vs. accumulated surface specific charge $q_\mathrm{A}$ yields the electrocapillarity coupling parameter $\varsigma$ with a value of $\varsigma=+0.657\pm0.007\,\mathrm{V}$. Overall, this independent measurement method yields a dependable value for the surface stress charge coefficient $\varsigma$. It represents a fundamental parameter for the material and can be utilized in the analysis of the actuation properties of the porous silicon samples or for other experiments where the electrochemistry of silicon is of importance.\\
	\section{Discussion}
	The microscopical cause of the electrochemical actuation phenomenon is rooted in surface stress, that is caused by the charging of the surface. Through our measurements it becomes apparent, that this relation is also valid for porous silicon. 
	
	The dependence of macroscopic dimensional change $\delta l/l$ on surface stress $f$ can be described by a model that assumes a porous material in the form of smooth, cylindrical, parallel oriented and hexagonally arranged pores, that stretch from top to bottom.\cite{Weissmueller2010} As the porous structure of porous silicon suits these assumptions, the model is thus appropriate to be utilized. The change in length $\delta l/l$ is then,\cite{Weissmueller2010}
	\begin{equation}
		\delta l/l=\delta \epsilon=-\frac{\alpha f}{3K}\frac{(1-\nu)}{(1-2\nu)},
		\label{Eq_pSi_strain}
	\end{equation}
	where $\alpha=1-\Phi$ denotes the solid volume fraction, $K=97.8\,\mathrm{GPa}$ the bulk modulus of silicon and $\nu=0.064$ Poisson's ratio for silicon in the $(110)$ direction.\cite{Hopcroft2010} Interestingly, the relation for $\delta l/l$ is similar to that of nanoporous gold, albeit their very different porous morphologies.\cite{Weissmueller2010} Based on Equation \ref{Eq_pSi_strain} and \ref{Eq_Maxwell} the already mentioned strain-charge coupling parameter $A^{*}$ reads
	\begin{equation}
		\label{Eq_A*}
		A^{*}=\frac{\mathrm{d}\varepsilon}{\mathrm{d}q_\mathrm{V}}=-\frac{\alpha\varsigma}{3K}\frac{(1-\nu)}{(1-2\nu)}.
	\end{equation}
	The value that has been determined for $\varsigma$ in this study suggests $A^{*}=-0.0015\,\mathrm{mm^3 C^{-1}}$. 
	In a next step this value, deduced from the measurement of the bulk silicon sample, is compared with the measurement of the porous silicon sample. Hence, the characteristic film stress charge coupling parameter $\xi$ is transformed into a strain charge coupling parameter. The necessary equation reads
	\begin{equation}
		\label{Eq_Hook}
		A^{*}=\frac{\varepsilon_\mathrm{||}}{q_\mathrm{V}}=\frac{1}{E_\mathrm{||}}\xi,
	\end{equation}   
	where $\varepsilon_\mathrm{||}$ denotes the strain and $E_\mathrm{||}$ the Young's modulus in the plane of the porous silicon layer, and $\xi = -296\,\mathrm{mV}$ for the measurement with perchloric acid. The crystal symmetry of the porous silicon layer can be approximated as in-plane, transverse isotropic, where the pores break the cubic crystallographic symmetry of the silicon pore walls, as a recent study finds.\cite{Thelen2021} Thus, $E_\mathrm{||}$ can be determined by
	\begin{equation}
		\label{Eq_E}
		\frac{1}{E_\mathrm{||}}=\frac{c_{33}}{-2 {c_{13}}^2+c_{33}\cdot(c_{11}+c_{12})}\approx\frac{1}{28.07\,\mathrm{GPa}},
	\end{equation}
	with $c_\mathrm{i}$ being the elastic coefficients\cite{Thelen2021,Lubarda2008} and approximating the difficult to measure coefficient $c_{12}$ as $c_{12}\le c_{13}$. Equation \ref{Eq_Hook} then yields $A^{*}=-0.011\mathrm{mm^3 C^{-1}}$ as a lower absolute limit, which corresponds to a strain amplitude by the porous silicon layer of $A^{*}\cdot q_\mathrm{V_{avg}}=1.69\pm0.04\cdot10^{-3}\%$ \revision{or approximately $50.5\,\mathrm{\upmu m}$ for the sample of length $l$}. The value $A^{*}=-0.011\mathrm{mm^3 C^{-1}}$ is about one order of magnitude smaller than $A^{*}=-0.0015\,\mathrm{mm^3 C^{-1}}$, the value obtained by Equation \ref{Eq_A*}. A reason for the present deviation could be due to the morphology of the porous structure of the here presented porous silicon. They deviate a lot from the idealized straight, round channel with a smooth surface, that underlies the theoretical calculations in Reference \cite{Weissmueller2010}. The SEM micrograph depicted in Figure S1 in the supplementary materials gives a distinct impression of the porous morphology. Thereby, the main pores are slightly meandering. Furthermore, smaller pores branch off the main pores. Thus, the pore walls are far from resembling their smooth equivalent assumed in the theoretical study. These side pores can be considered as an extreme increase in roughness. For silicon, surprisingly, such an increase in roughness even results in an enhanced electrochemo-mechanical coupling of the surface stress induced actuation. This is in stark contrast to metals with large Poisson ratio such as copper or gold.\cite{Weissmuller2008}\\
	For nanoporous gold, a well investigated material in the regard of electrochemical actuation, a strain-charge coupling parameter of $A^{*}=0.0475 \pm 0.0010\,\mathrm{mm^3C^{-1}}$ has been found.\cite{Stenner2016} Thus, the absolute value of $A^{*}=-0.011\mathrm{mm^3 C^{-1}}$ obtained in this study for porous silicon lies on the same order of magnitude, albeit being approximately 4 times smaller. Aside from the entirely different porous structure of nanoporous gold, a distinct difference between the materials is their conductive properties. Whereas gold is an excellent conductor, silicon is semi-conducting. In gold the electrons, that counter the ionic charge carriers which accumulate in the Helmholtz layer on the electrolyte side of the interface, assemble in an $\buildrel _\circ \over {\mathrm{A}}$ thin layer beneath the interface. Conversely, when an oxidized silicon wall is in contact with an electrolyte solution, a space charge region beneath the silicon-oxide interface emerges. The charge carriers, holes in the case of p-doped silicon, are distributed over the width of the space charge region.\cite{Zhang2007} The width $w$ of the space charge region can be calculated by
	\begin{equation}
		\label{Eq_space-charge}
		w=\sqrt{\frac{2\varepsilon_\mathrm{s}}{e N_\mathrm{D}} (V_\mathrm{fb}+\frac{k_\mathrm{B}T}{e})},
	\end{equation}
	where $\varepsilon_\mathrm{s}\,=\,\varepsilon_\mathrm{0}\cdot\varepsilon_\mathrm{r}$, with $\varepsilon_\mathrm{0}$ and $\varepsilon_\mathrm{r}$ denoting the vacuum- and relative permittivity.\cite{Zhang2007} $N_\mathrm{D}$ is the doping concentration, $V_\mathrm{fb}$ the flat band potential, where the bands at the electrolyte interface are not bend but straight and $k_\mathrm{B}$ and $T$ denote the Boltzman constant and the temperature. It can be readily seen, that the width of the space charge region foremost depends on the doping concentration $N_\mathrm{D}$ and the flat band potential $V_\mathrm{fb}$. With the assumption of a pure silicon interface, thus $\varepsilon_\mathrm{r}=11.68$, a doping concentration of $N_\mathrm{D}=5\cdot10^{18}\,\mathrm{cm^{-3}}$ and a flat band potential of $0.4\,\mathrm{V}$, the space charge region has a width of $w=9.8\,\mathrm{nm}$.\cite{Gray2001,Lehmann2002,Searson1991} Thus, the charge carriers that cause the surface stress are distributed in a layer about one order of magnitude thicker than in gold. Hence, the actuation response upon an equal accumulated charge is possibly smaller, which is, what is observed here.\\
	Overall in this study, we showed the versatile nature of porous silicon with regard to possible electrochemical actuation applications. The investigation of the characteristic electrochemical behavior in acidic- and salt solutions and the thereby induced surface stress related actuation lead to a deeper understanding of porous silicon. Furthermore, surprisingly we find an inversely proportional relation of the actuation on potential and charge. For the first time, the vital, fundamental surface stress charge coupling parameter $\varsigma$ could be determined for silicon surfaces.\\
	Similarly as it has been demonstrated for liquid-infused porous structures with regard to confinement-controlled phase selection \cite{Henschel2007,Huber2015}, adaptable surface wetting \cite{Seker2008WettingEvaporation,Xue2014}, topography \cite{Wang2018}, photonic \cite{Sentker2018,Sentker2019} and mechanical properties \cite{Style2021SolidLiquidMaterials,Zhang2021LiquidBasedMaterials} our study indicates that electrolyte-infused nanoporous solids allow for a quite simple fabrication of materials with integrated electro-actuorics.\\
	Finally, our study is also a fine example, how the combination of soft and hard matter opens up the possibility of using multi-physical couplings in geometrical confinement from the nano- via the meso- to the macroscale to design robust hybrid materials with integrated functionality, as it can be found in many biological composites \cite{Eder2018,Gang2020SoftMatterBook}.
	\section{Experimental}
\begin{figure}
		\centering
		\includegraphics{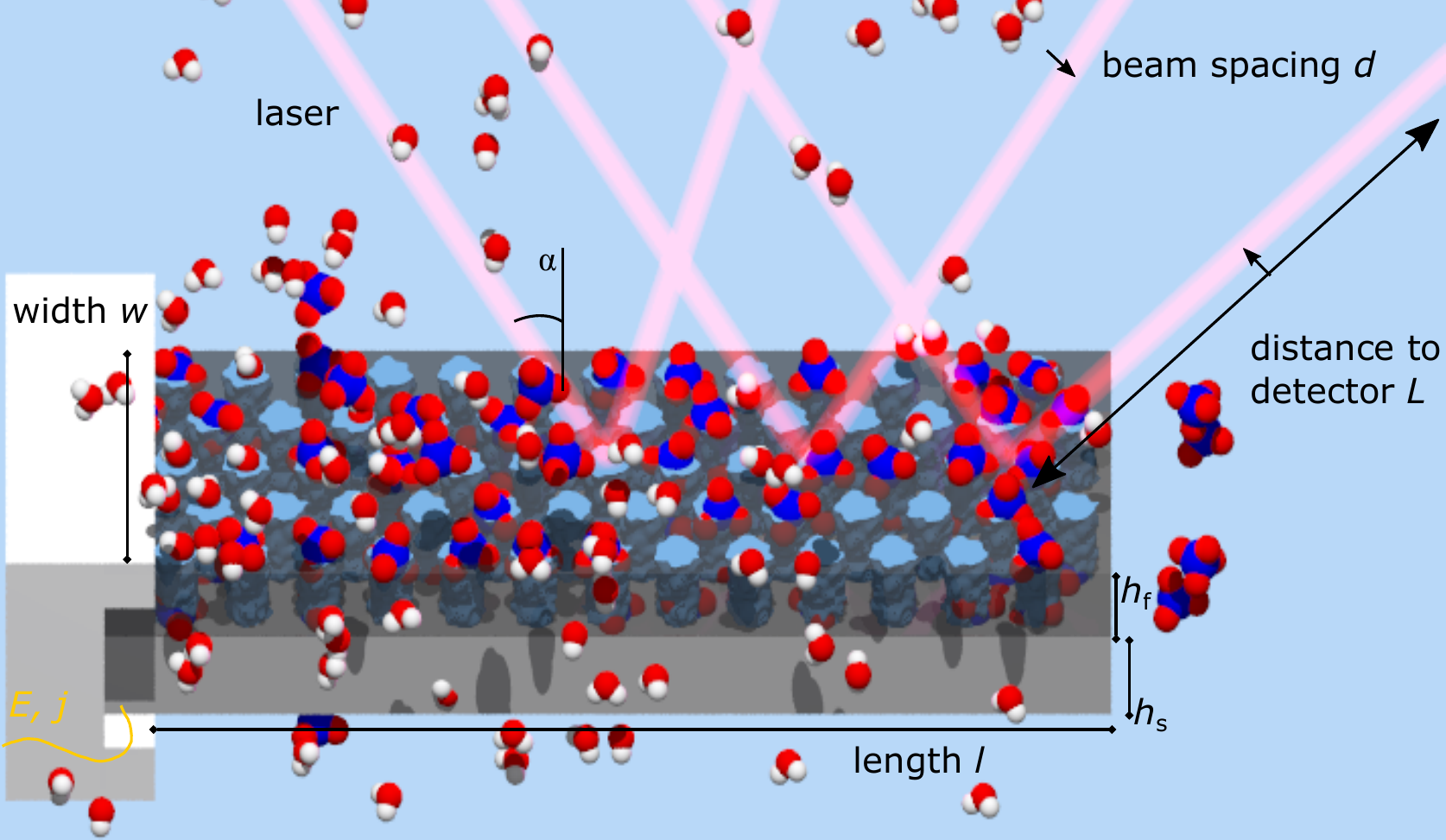}
		\caption{\textbf{Schematic depiction of the sample geometry and measurement setup.} The sample consists of a porous layer (dark gray) attached to a remaining bulk silicon layer on the bottom (light gray). The pores are filled with electrolyte solution indicated by the blue color and the $\mathrm{HClO}_4$ (blue, red) and $\mathrm{H}_2\mathrm{O}$ (red, white) molecules. The dimensions of the fabricated sample are length $l$, width $w$ and the thickness of the substrate $h_\mathrm{s}$ and the porous silicon film $h_\mathrm{f}$ respectively. Moreover, the figure depicts the experimental cantilever bending setup. The sample acts as the working electrode. It is contacted by an aluminum contact \revision{and an attached gold wire on the left side to apply potential $E$ and measure current $j$.} The potential $E$ in the electrolyte solution can be changed, so that the $\mathrm{ClO}_4^{-}$ anions are accumulated on the bulk silicon surface, or the porous silicon pore surface respectively, or so that the anions are expelled. The result is an induced surface stress which leads to a contraction of the sample followed by its subsequent expansion. Since the underlying bulk silicon substrate clamps the film, the film is stressed which leads to a curvature of the substrate in return. A laser beam is split into an array and gets reflected into a CCD detector, to measure the curvature $\kappa$ of the sample. \revision{The laser here is aimed at the top side of the sample for illustrating purposes, whereas in the actual experimental section it is reflected off the backside. The laser array is also aimed at the fare end of the clamped side, to avoid an influence on the bending by the clamping.} The distance between sample and detector is denoted by $L$, the incident angle of the laser beam array by $\alpha$ and the spacing between the individual laser beams hitting the detector by $d$. The sample is clamped on the left by PTFE. The whole setups is housed in an electrochemical cell, that is fabricated out of PTFE as well and sealed on the top with an optically transparent glass plate.}
		\label{Fig_Setup}
	\end{figure}

	\noindent Two types of sample are investigated in this study -- bulk silicon and porous silicon. The base material for the fabrication of both types of samples is p-doped single crystalline silicon. The supplier (Si-Mat Silicon Materials GmbH) gives a thickness of $100\pm10\,\mathrm{\upmu m}$ for the wafers, a (100) orientation and a resistivity of $0.01\,-\,0.02\,\mathrm{\Omega cm}$. Both back- and frontside of the wafer are polished. The electrochemo-mechanical actuation properties of bulk silicon, particularly the response of surface stress $f$ to accumulated charge $q$, i.e. the electrocapillarity coupling parameter $\varsigma$, are investigated on this same base material. The sample for such a measurement is prepared in the following way. A p-doped wafer is oxidized in an oven at $850\,\mathrm{^\circ C}$ for 16 hours, with a heating and cooling ramp of $200\,\mathrm{^\circ C h^{-1}}$, to create a thick, non-conductive insulation layer of silicon dioxide.\cite{Canham2015} The thickness of this layer is determined to be $96.3\,\mathrm{nm}$ by ellipsometry. Subsequently, the insulation layer is removed on the frontside by a hydrofluoric acid (HF) dip in $10\,\%$ aqueous HF solution for 5 minutes. The actual sample is then prepared from the fabricated silicon material by cleaving.\cite{Sailor2011} The sample is shaped rectangularly with a long side of $5.37\mathrm{cm}$ and a short side of $1.49\,\mathrm{cm}$. The surface area which is in contact to the solution, though, is smaller. Not all of the sample is immersed into the electrolyte solution, as the top of the sample is clamped by Polytetrafluorethylen (PTFE). The surface area in contact to the electrolyte is determined precisely in section \ref{sec_bulk} and amounts to $6.38\,\mathrm{cm^2}$.\\
	Porous silicon is prepared by etching the frontside of the silicon wafer with the prepared backside insulation layer in an electrochemical procedure with HF. The wafer is electrically contacted by aluminum foil and mounted into an electrochemical cell. The cell is made from PTFE, to safely handle HF. A fluoro-elastomer-made O-ring with an inner radius of $3.30\,\mathrm{cm}$ seals the contact between wafer and cell. Firstly, the insulation layer is removed with the above described HF-dip. Then the electrolytic solution is exchanged and the cell is filled with a 2:3 volumetric mixture of HF (48\%, Merck Emsure) and ethanol (absolute, Merck Emsure). The HF electrolyte is allowed to equilibrate for 5 minutes. A platinum counter electrode (CE) is inserted into the cell from the top. It acts as the cathode, whereas the silicon is the anode. The accordingly called anodization or etching process starts by applying a constant current of $428\,\mathrm{mA}$ which equals a current density of $12.5\,\mathrm{mA cm^{-2}}$. After 1 minute the current is switched off and the HF solution is removed from the cell. In a final step the resulting porous silicon is rinsed three times with distilled water, demounted from the cell and dried at ambient conditions for 3 hours. The actual sample is prepared from the fabricated porous silicon material equally to the bulk silicon sample by cleaving. The sample has a length of $l=2.990\pm0.002\,\mathrm{cm}$ and a width of $w=1.796\pm0.002\,\mathrm{cm}$. To stress this point, the porous layer is not detached in a so called electropolishing step from the remaining bulk silicon. Rather the porous layer stays attached to the bulk silicon. Figure \ref{Fig_Setup} gives a sketch of the sample and its dimensions.\\
	Transmission electron microscopy is used to record a micrograph of a cross section of the porous silicon (FEI Helios G3 UC). To measure the thickness of the top and bottom porous layer, a profile is recorded with a scanning electron microscope (Zeiss Leo 1530). Further properties of the resulting porous silicon as mean pore diameter, porosity and internal surface area are characterized in a nitrogen sorption isotherm setup (Quantacrome autosorb iQ).\\  
	A simple and sensitive experiment for the measurement of the actuation properties represents an in-situ cantilever bending investigation, described in great detail by Smetanin et al.\cite{Smetanin2008} The respective sample, bulk- or porous silicon, is installed into the in-situ cantilever bending setup. It is electrically contacted at the top end by an aluminum contact and an attached gold wire. As the sample is exposed to an acidic solution, the contact needs to be protected from corrosion. A scheme of the setup can be found in Figure \ref{Fig_Setup}.
	The sample is immersed in one of two aqueous electrolytic solutions -- perchloric acid ($\mathrm{HClO}_4$) with a concentration of $1\,\mathrm{mol\, l^{-1}}$ or a sodium chloride ($\mathrm{NaCl}$) solution with a concentration of $154\,\mathrm{mol\, l^{-1}}$ (isotonic saline solution).  $\mathrm{HClO}_4$ is prepared from 70\% perchloric acid (Merck Suprapur) and deionized water ($18.2\,\mathrm{M\Omega}$). Deionized water is used as well to solve $\mathrm{NaCl}$ (Merck Suprapur). Before a change of electrolyte solution, the sample is rinsed three times with deionized water and dried at ambient conditions for three hours. The surface area of the porous silicon sample, which is in contact to the solution amounts $5.37\,\mathrm{cm^2}$. The contact surface of the bulk silicon sample is determined in the results chapter. A carbon cloth CE and a reversible hydrogen reference Electrode (RE, Gaskatel HydroFlex) are inserted into the chosen solution. All electrochemical potentials in this work are denoted versus the standard hydrogen electrode (SHE). The electrodes and the sample are connected to a potentiostat (Metrohm-Autolab PGSTAT 30) and thus electrochemical measurements can be performed.\\
	A potential $E$ is applied between sample and CE, which is measured and controlled via the RE by the potentiostat. 
	Additionally, the potentiostat measures the current $j$ that flows between sample and CE and provides the accordingly consumed charge $Q$ as well. The applied potential can be changed with a linearly varying slope between a lower and higher reversal potential, a so-called cyclo voltammetry measurement. By these means, certain electrochemical characteristics as the capacitance of the sample or the presence of electrochemical reactions with an electron exchange, i.e. reduction or oxidation reactions, can be inferred. The rate $\mathrm{d}E/\mathrm{d}t$ with which the potential is varied is called the sweep- or scan rate. The theoretical potential range, in which a CV can be conducted in aqueous solutions, lies in between $0\,\mathrm{V}$ and $1.223\,\mathrm{V}$. Below, a reduction of the soluble, i.e. water, occurs and hydrogen gas forms. Above, at potential values higher than $1.223\,\mathrm{V}$, water itself starts to be oxidized and oxygen gas emerges, although in practice different electrolyte solutions have an overpotential.\\
	Another type of measurement used is an instant change of a lower potential to a higher value and vice versa in a step-like fashion, also known as step-coulombmetry. After changing the potential it is subsequently kept constant for a period of time, resembling a square potential. In this way, it is possible to determine the kinetics of the electrochemical processes and the accompanying change in actuation. Furthermore, the characteristic time constant of the evolution in charge and the cantilever bending signal upon this instant change in potential can be determined. Lastly, the potential can also be kept constant at a single value while the current is recorded.\\ 
	Altogether, when the potential is changed, the solution's ions are accumulated on the bulk silicon surface or on the surface of the porous silicon layer respectively. Thus, charge carriers are not accumulated on and dispersed off the wafer's insulated backside but solely on the frontside surface. The accumulated charge thereby changes accordingly the surface stress. This leads to a bending of the wafer. It can be detected and measured by a laser setup, where a laser beam array is reflected off the backside of the cantilever-like sample and detected in a CCD sensor. \revision{The setup (Multi-Optical Stress Sensor, k-Space\cite{Floro1995,Floro1997}) is described in great detail in Reference \cite{Smetanin2008}. To minimize the impact of vibrations the setups is installed on a vibration isolated table. The setup is surrounded by isolation housing, to reduced vibrations by air flow. Thermal drift is reduced by air conditioning the room at $21\,\mathrm{^\circ C}$. Reference \cite{Roschning2019} determined a resolution of bending radius of $250\,\mathrm{km}$ for this specific setup, which is sufficient for the measurements conducted here.} The time resolution is $1\,\mathrm{s}$. The spot spacing $\Delta d$ of the array of laser beams changes with the curvature of the sample and thus the evolving surface stress. Thereby, the resulting resolved curvature change parallel to the long axis of the wafer $\Delta\kappa$ is the relevant parameter and is determined by,\cite{Smetanin2008}
	\begin{equation}
		\label{Eq_curvature}
		\Delta\kappa = \frac{\Delta d}{d_0}\frac{\cos(\alpha)}{2L n},
	\end{equation}  
	where $d_0$ is the unimpeded spacing averaged over 1 minute, $\alpha=4\,^\circ$ the incident angle of the beam array, $L=110\,\mathrm{cm}$ the distance from the laser to the CCD camera and $n=1.33$ the refractive index of water as a good approximation for the diluted aqueous electrolyte solution. In case of the bulk silicon, the curvature change is a result of a change in the surface stress and is related to the curvature via Stoney's equation,\cite{Stoney1909,Doerner1988,Smetanin2008, Dittrich2018MasterThesis}  
	\begin{equation}
		\label{Eq_Stoney}
		\Delta f=-\frac{1}{6}M h^2 \Delta\kappa,
	\end{equation}
	where $M=180\,\mathrm{GPa}$ is the biaxial modulus\cite{Hopcroft2010} and $h=106.71\pm1.03\,\mathrm{\upmu m}$ is the thickness of the bulk silicon sample. For the porous silicon sample a film stress $\sigma$ evolves in the porous silicon film and assuming it is isotropic in the film plane, Stoney's equation relates it to $\Delta\kappa$ via
	\begin{equation}
		\label{Eq_Stoney_Film}
		\Delta \sigma=-\frac{1}{6}M\frac{ h_\mathrm{S}^2}{h_\mathrm{f}} \Delta\kappa.
	\end{equation}
	Thereby, $h_\mathrm{S}$ and $h_\mathrm{f}$ denote the thickness of the substrate and the film respectively and are given in the results section. \revision{It is important to notice that the analysis of the film stress via Stoney's Equation \ref{Eq_Stoney_Film} only demands knowledge of the mechanics of the substrate and not the film. This assumption that the contribution of the film to the overall bending stiffness can be ignored requires, that the film's thickness $h_\mathrm{f}$ does not greatly exceed the substrates thickness by a factor of $2\cdot10^{-3}$.\cite{Nix1989} Here, $h_\mathrm{f}/h_\mathrm{S}\approx5.8\cdot10^{-3}$ is well within that range.}\\
	%
	%
	%
	%
	%
	%
	%
	%
	\medskip
	\textbf{Supporting Information} \par 
	Supporting Information is available from the Wiley Online Library or from the author.

	\medskip
	\textbf{Acknowledgements} \par 
	We acknowledge the excellent preliminary experiments conducted by Guido Dittrich (Hamburg University of Technology) during his master thesis. We also acknowledge the excellent support regarding the scanning- and transmission electron microscopy from Tobias Krekeler (Hamburg University of Technology) as well as the beam bending measurements by Benedikt Roschning (Hamburg University of Technology), and fruitful discussions with J\"org Weissm\"uller (Hamburg University of Technology and Helmholtz-Zentrum Hereon) and Norbert Huber (Helmholtz-Zentrum Hereon and Hamburg University of Technology).\\ 
	This work was supported by the Deutsche Forschungsgemeinschaft (DFG) within the Collaborative Research Initiative SFB 986 "Tailor-Made Multi-Scale Materials Systems" Projectnumber 192346071. We also acknowledge the scientific exchange and support of the Centre for Molecular Water Science CMWS, Hamburg.\\
	All data that is needed to evaluate the conclusions is presented in the paper and in the supplementary materials. The raw data of the electrochemical actuation experiments are available at TORE (https://tore.tuhh.de/), the Open Research Repository of Hamburg University of Technology, at the doi: .\\
	M.B. and P.H. conceived the experiments. M.B. performed the material synthesis and the actuator measurements. M.B. and P.H. evaluated the data. M.B. and P.H. wrote and proofread the manuscript.\\ 
	The authors declare that they have no competing interests.\\
	
	\medskip
	\clearpage

\end{document}